\documentclass{aa}  

\usepackage{natbib}	
\usepackage{graphicx}
\usepackage{txfonts,epsfig,graphicx,url,twoopt}
\usepackage[breaklinks=true]{hyperref} 
\hypersetup{colorlinks=true,citecolor=blue,linkcolor=blue}
\bibpunct{(}{)}{;}{a}{}{,}     
\usepackage[usenames,dvipsnames]{color}

\definecolor{dgreen}{rgb}{0.0,0.5,0}

\usepackage{xcolor}

\begin{document}
   \title{Modeling dust growth in protoplanetary disks: The breakthrough case}

   \author{J. Dr\k{a}\.{z}kowska\inst{1}\fnmsep\inst{2}
          \and
          F. Windmark\inst{1}\fnmsep\inst{2}
          \and
          C.P. Dullemond\inst{1}
          }

   \institute{Heidelberg University, Center for Astronomy, Institute for Theoretical Astrophysics, Albert-Ueberle-Str.\ 2, 69120 Heidelberg, Germany\\
         \email{drazkowska@uni-heidelberg.de}
         \and
         {Member of the International Max Planck Research School for Astronomy and Cosmic Physics at the Heidelberg University}
             }

   \date{Received 25 February 2014 \slash Accepted 20 May 2014}


 
  \abstract
   {
   Dust coagulation in protoplanetary disks is one of the initial steps toward planet formation. Simple toy models are often not sufficient to cover the complexity of the coagulation process, and a number of numerical approaches are therefore used, among which integration of the Smoluchowski equation and various versions of Monte Carlo algorithm are the most popular.
   }
   {
   Recent progress in understanding the processes involved in dust coagulation have caused a need for benchmarking and comparison of various physical aspects of the coagulation process. In this paper, we directly compare the Smoluchowski and Monte Carlo approaches to show their advantages and disadvantages.
   }
   {
   We focus on the mechanism of planetesimal formation via sweep-up growth, which is a new and important aspect of the current planet formation theory. We use realistic test cases that implement a distribution in dust collision velocities. This allows a single collision between two grains to have a wide range of possible outcomes but also requires a very high numerical accuracy.
   } 
   {
   For most coagulation problems, we find a general agreement between the two approaches. 
   However, for the sweep-up growth driven by the "lucky" breakthrough mechanism, the methods exhibit very different resolution dependencies. With too few mass bins, the Smoluchowski algorithm tends to overestimate the growth rate and the probability of breakthrough. The Monte Carlo method is less dependent on the number of particles in the growth timescale aspect but tends to underestimate the breakthrough chance due to its limited dynamic mass range.}
   {We find that the Smoluchowski approach, which is generally better for the breakthrough studies, is sensitive to low mass resolutions in the high-mass, low-number tail that is important in this scenario. To study the low number density features, a new modulation function has to be introduced to the interaction probabilities. As the minimum resolution needed for breakthrough studies depends strongly on setup, verification has to be performed on a case by case basis.
   }


   \maketitle

\section{Introduction}

Dust coagulation is an important astrophysical process, particularly in the case of planet formation. To form a planet, micron-sized dust particles have to collide and stick to form ever larger dust aggregates. The main problem with this picture is what is known as the "meter-size barrier": once particles reach a size of decimeter to meter, they acquire relative velocities that are too large to allow hit-and-stick growth. Instead, collisions at these speeds, which are on the order of tens of meters per second, lead to the destruction of the colliding aggregates. This is often called the "fragmentation barrier". Moreover, already at sizes of a millimeter, we encounter the "nonsticking problem", which is often called the "bouncing barrier": while the aggregates do not fragment upon collision at these sizes, they do not stick together either, meaning that the growth stalls. All these barriers are posing a major problem in our understanding of planet formation. In recent years, however, several new ideas have appeared in the literature to overcome these problems. 

For example, the sweep-up growth scenario has been proposed as a solution for the growth barriers issue \citep{2012A&A...540A..73W}. In this scenario, a few boulders can grow large by colliding with numerous smaller pebbles through the fragmentation with mass transfer process \citep{2005Icar..178..253W,2009MNRAS.393.1584T,2013A&A...559A.123M}. The growth of the pebbles is suppressed due to bouncing and/or fragmentation between themselves, while a few larger aggregates, or seeds, might form if the distribution of impact velocities due to stochastic turbulence is taken into account \citep{2012A&A...544L..16W, 2013ApJ...764..146G,2013ApJ...774L...4M}. Thanks to the low velocity tail of the distribution, some grains can be "lucky" enough to not experience any destructive collisions and undergo only low-velocity sticking collisions, breaking through the growth barriers. This scenario enables the formation of planetary embryos while still keeping the disk dusty, which is consistent with observations.

Key to the trustworthiness of the conclusions derived from numerical models is the reliability of the codes and algorithms used. The problem of coagulation is extremely complex and nonlinear, and with the exception of some very simple coagulation kernels, no analytic solutions exist. So how do we know if the results of our codes are indeed correct? One way is to treat the problem with at least two distinct methods and compare the results.

Over the years, different approaches have been developed to study the dust coagulation problem. Besides numerous semi-analytic models, two main numerical approaches are used nowadays: direct numerical integration of the Smoluchowski equation and various Monte Carlo codes. The former is traditional approach, which has been used in different versions by \citet{1980Icar...44..172W, 1981Icar...45..517N, 1989Icar...77..330W, 1990Icar...83..205O, 2005A&A...434..971D, 2005ApJ...625..414T, 2006ApJ...640.1099N, 2008A&A...480..859B, 2009ApJ...707.1247O, 2010A&A...513A..79B, 2012ApJ...753..119C} and many others. 
This approach is often used when comparing dust coagulation models to observations, as it allows us to model the dust evolution in the global disk over very long timescales.
The Monte Carlo approach is based on work by \citet{1975MNRAS.170..541G} and is used in one form by \cite{2007A&A...461..215O} and in another by \citet{2008A&A...489..931Z}, subsequently used by \citet{2012A&A...537A.125J,2013A&A...552A.137R} and \citet{2013A&A...556A..37D}. The Monte Carlo approach is useful to test different coagulation models and to include different properties of dust particles, such as the internal grain structure. It is also better suited to use along with hydrodynamic grid codes.

The two methods are usually benchmarked using analytical solutions of the coagulation equation that are available for three idealized growth kernels (see, e.g., \citealt{1990Icar...83..205O}, \citealt{1990Icar...88..336W}, and \citealt{2000Icar..143...74L}). However, these kernels do not necessarily represent any realistic growth scenario in the protoplanetary disk. In this work, we perform an explicit comparison between the two approaches for the first time. In this comparison, we focus on the sweep-up growth scenario, which is challenging to model for both of the methods. In particular, it was already asserted by \citet{2012A&A...548C...1W} that an artificial breakthrough may occur when a low mass resolution is used in the Smoluchowski method. We study this issue in more detail, and we show that not only the high resolution but also a careful treatment of interactions in low particle number density bins is needed to avoid the nonphysical growth.

Until now, the sweep-up growth triggered by the "lucky" growth was modeled using the Smoluchowski code only. Using a two-dimensional Monte Carlo code, \citet{2013A&A...556A..37D} showed that sweep-up growth can occur at the inner edge of dead zone, but it was triggered by radial transport of big bodies grown in a dead zone in this case. In this paper, we implement the relative velocity distribution in the Monte Carlo code and directly compare results of the two approaches. We also present some of their major features and differences.

This paper is organized as follows: we describe both of our numerical models in Sect.\ \ref{sub:model}. In Sect.\ \ref{sub:comp}, we compare results obtained with both codes. We discuss issues related to numerical convergence of both methods in Sect.\ \ref{sub:res}. We summarize our findings in Sect.\ \ref{sub:last}.

\section{The numerical models}\label{sub:model}

We study the Smoluchowski approach using the code developed by \citet{2008A&A...480..859B} and \citet{2010A&A...513A..79B}, along with an impact velocity distribution implemented as described in \citet{2012A&A...544L..16W}. In this code, we let the dust-grain number density $n(m,r,z)$ be a function of the grain mass $m$, the distance to the star $r$, and the height above the midplane $z$, and give it in number of particles per unit volume per unit mass. The dust evolution can then be solved by integration.

However, discretization of the problem is necessary in the integration process, which can lead to a significant numerical diffusion in mass-space, because having a finite number of grid points means that particle collisions do not necessarily lead to particle masses $m_{\rm p}$ that directly correspond to one of the logarithmically spaced sampling points. The approach by \citet{2008A&A...480..859B} was to implement an algorithm that distributes the mass of the resulting particle into two adjacent mass bins corresponding to grid points ${\rm i}$ and ${\rm i}+1$, $m_{\rm i} < m_{\rm p} < m_{\rm i+1}$, according to
\begin{equation}\label{epsilon}
	\epsilon = \frac{m_{\rm p} - m_{\rm i}}{m_{{\rm i}+1}-m_{\rm i}},
\end{equation}
where $\epsilon \cdot m_{\rm p}$ is put into mass bin $m_{\rm i+1}$, and $(1-\epsilon)\cdot m_{\rm p}$ is put into mass bin $m_{\rm i}$. This algorithm is based on the work of \citet{1969JAtS...26.1060K} and is adopted by most of the modern dust coagulation codes. This approach, however, leads to some numerical diffusion, as $m_{\rm i+1} > m_{\rm p}$ means that mass is inserted into a mass bin that corresponds to a larger mass than the mass of physical particle created in the collision. If the spacing between the mass bins is too coarse, this leads to a significant, artificial growth rate speed-up \citep{1990Icar...83..205O}. We show that the same effect strongly affects the number of seeds formed in the "lucky growth" scenario; however, the problem is more severe and requires a careful approach to low number density regions in this case.

In the Monte Carlo method, we use the representative particles approach described by \citet{2008A&A...489..931Z} and implemented by \citet{2013A&A...556A..37D}. We have slightly modified the method to match the vertical treatment of the Smoluchowski code (described by \citealt{2008A&A...480..859B}). Instead of performing an explicit vertical advection, we redistribute the particles according to a Gaussian distribution with a width:
\begin{equation}\label{hdust}
H_{\rm{dust}} = H_{\rm{gas}} \left[ 1 + \frac{\min{\left(0.5,\rm{St}\right)\left(1+\rm{St}^2\right)}}{\alpha}\right]^{-\frac{1}{2}},
\end{equation}
where $H_{\rm{gas}}$ is the pressure scale height of the gas, $\rm{St}$ is the particle's Stokes number and $\alpha$ is the turbulence strength parameter. In this way, we account for the reduction in the collision rate between small and large grains due to their different vertical settling. This occurs because small particles are more strongly affected by the turbulent diffusion, and, thus, their density in the midplane of the disk is lower than in the case of large particles.

One of the assumptions of the representative particle approach that we implement is that one representative particle, or swarm, represents a constant amount of mass, which is equal to
\begin{equation}\label{mswarm}
M_{\rm{swarm}} = M_{\rm{tot}} / N_{\rm{swarms}},
\end{equation}
where $M_{\rm{tot}}$ is the total mass of dust present in the computational domain and $N_{\rm{swarms}}$ is the number of swarms used. 
In other words, the total mass of dust $M_{\rm{tot}}$ is divided into $N_{\rm{swarms}}$ equal-mass units. Each of these units represents physical particles of mass $m_i$ ($m_i \ll M_{\rm{swarm}}$), but the number of these particles $N_i$ has to be such that $N_i \cdot m_i = M_{\rm{swarm}}$. The algorithm fails, if, for example, a physical coagulation kernel would lead to formation of only one massive particle with mass $m_{\rm p}$, while keeping all the other particles small. If $m_{\rm p}<M_{\rm{swarm}}$, the single big particle cannot be resolved, because it involves less mass than the smallest available unit $M_{\rm{swarm}}$.  
Increasing the number of swarms, $N_{\rm{swarms}}$, lowers $M_{\rm{swarm}}$ and thus improves the mass resolution of the method; however, the computation time increases quadratically with the number of swarms used. The dynamic mass range of the representative particle approach is limited by the number of swarms used. This issue can be overcome by implementing a more advanced algorithm, as the “distribution method” proposed by \citet{2008ApJ...684.1291O} that involves continuously adjusting $M_{\rm{swarm}}$ by splitting and merging the swarms. This method was later used in the context of accretion among planetesimals \citep{2010Icar..210..507O}, allowing the Monte Carlo method to resolve a runaway coagulation kernel. However, the method was not yet tested with velocity distributions or complicated collision models that are needed to break through the growth barriers.

\section{Comparison of results obtained with both methods}\label{sub:comp}

Because our Monte Carlo method is not capable of the same dynamic mass range as the Smoluchowski method to directly compare the two methods, we choose a setup where the particle breakthrough (i.e.\ where the "lucky" particles can start to grow by mass transfer) occurs at relatively high particle number density, which is possible to resolve with our representative particle approach (see the previous section).

For the disk model, we use the minimum-mass extrasolar nebula \citep{2013MNRAS.431.3444C} at 1\ AU. The gas surface density is $\Sigma_{\rm{gas}}=9900$~g~cm$^{-2}$, and the temperature is $T=280$~K. We assume a turbulence of $\alpha=10^{-2}$ and a standard dust to gas ratio of $10^{-2}$. We take a relative velocity between the dust grains driven by Brownian motion and turbulence into account, by calculating the root-mean-square impact velocity $v_{\rm{rms}}$ from the formulas derived by \citet{2007A&A...466..413O}, and we assume a Maxwellian distribution of the impact velocity.

We consider sticking, fragmentation, and mass transfer as possible outcomes of collision, which we refer to as the SF$+$MT model \citep{2010A&A...513A..56G, 2012A&A...544L..16W}. The collision outcome is determined by taking the sticking and fragmentation$\slash$mass transfer probabilities $P_{\rm{s}}(v_{\rm{rms}})$ and $P_{\rm{f\slash mt}}(v_{\rm{rms}})$ into account, which are calculated analogically as in \citet{2012A&A...544L..16W}. We take the fragmentation threshold velocity to be $v_{\rm{f}}=50$ cm~s$^{-1}$. If a collision should lead to fragmentation but the mass ratio between the colliding particles mass is $m_1/m_2>20$ ($m_1 > m_2$), we assume projectile fragmentation with mass transfer with a 10$\%$ efficiency; that is the larger particle gains 10$\%$ of mass of the smaller one during the event. Realistic values of the collision parameters are poorly constrained, as discussed in Sect.~\ref{sub:last}.

Implementing the same setup in both of the codes, we perform a number of runs by varying the numerical resolution by around one order of magnitude. In the Smoluchowski code, we use from 3 to 40 mass bins per decade. In the Monte Carlo code, we use from 12,000 to 120,000 representative particles, and we repeat each run ten times with different random seeds. 

The Monte Carlo method relies on random numbers used to determine which particles are participating in the subsequent collisions and to calculate collision time steps \citep{2008A&A...489..931Z}. Thus, outcomes of the Monte Carlo runs performed with different random seeds vary despite using the same setup. In the velocity distribution case, this effect is even stronger, meaning that multiple runs with different random seeds are necessary. This causes our high resolution Monte Carlo models to need a few days on an 8 core 3.1 GHz AMD machine. For comparison, the Smoluchowski models with 40 bins per mass decade take about of one hour on a single core processor.

\begin{figure}
   \centering
   \includegraphics[width=\hsize]{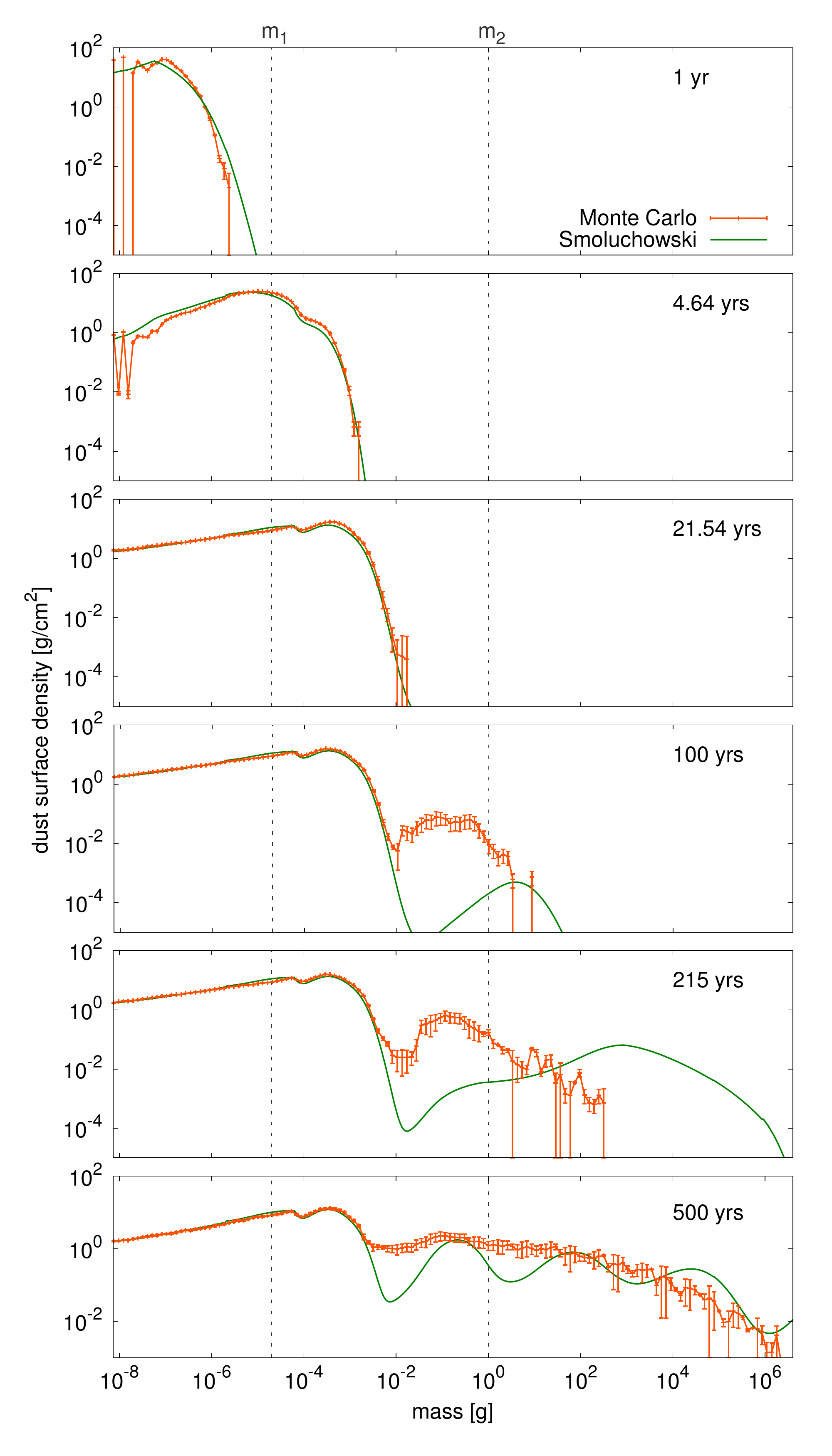}
      \caption{Comparison of the mass distribution evolution obtained using our Smoluchowski and Monte Carlo codes. In a standard case when an insurmountable fragmentation barrier is present, the two methods perfectly agree (three upper panels); however, we encounter some differences between the results obtained with the two approaches (three bottom panels) with the possibility of breakthrough. For the Smoluchowski code, the presented results were obtained with a resolution of 30 mass bins per decade. For the Monte Carlo code, the results are averaged from the simulations using $120,000$ particles. The “check points”, m$_1$ and m$_2$, which are indicated with the dotted lines, are used to quantitatively compare dust growth timescales in Sect.\ \ref{sub:timescales}.}
         \label{fig:SFMT}
\end{figure}

Figure~\ref{fig:SFMT} shows the time evolution of the dust mass distribution obtained with both of the codes. For the Smoluchowski code, we display results obtained with the resolution of 30 bins per mass decade. For the Monte Carlo code, we display averaged results, along with their scatter, of our highest resolution simulations with 120,000 particles, which additionally showed a breakthrough before 100 yrs; otherwise, the figure gets indecipherable due to high noise.

The early evolution, as seen in the upper panels of Fig.\ \ref{fig:SFMT}, is very similar in both codes, and the point at which the dust distribution hits the fragmentation barrier, where the growth of the majority of particles is hindered due to fragmentation that occurs when the impact velocities exceed the fragmentation threshold velocity $v_{\rm{f}}$, is identical ($m \cong 10^{-2}$~g, $t \cong 20$~yrs). 
If the fragmentation with mass transfer collisions are not included and there is no possibility of growth beyond the fragmentation barrier, the steady state is represented by the third panel of Fig.\ \ref{fig:SFMT}, and the two methods agree perfectly.
However, as we include the breakthrough possibility, we encounter some differences between the two approaches, which are visible at later stages of the evolution. Both of the codes reveal that some of the particles can grow beyond the fragmentation barrier thanks to the low-velocity sticking collisions. The growth is very quick in general and meter-sized bodies are formed within $<$1000 yrs. However, the breakthrough generally occurs later in the Monte Carlo code, and the population of big grains lacks the characteristic waves seen in the distribution obtained from the Smoluchowski code, which can be seen in the bottom panel of Fig.\ \ref{fig:SFMT}. The differences in the late stages of evolution are caused by the restricted dynamic mass range of our representative particle approach that was discussed in Sect.\ \ref{sub:model}. We further discuss issues related to resolution dependencies of both methods in the following section.

\section{Resolution dependence}\label{sub:res}

The results obtained with both codes exhibit resolution dependence. In this section, we discuss specific issues connected to the numerical convergence of both Smoluchowski and Monte Carlo methods.

\subsection{Growth timescales}\label{sub:timescales}

To quantitatively compare dust growth obtained in both codes, we establish “check points” m$_1$ and m$_2$, as marked with the dotted lines in Fig.\ \ref{fig:SFMT}. We arbitrarily choose m$_1 = 2\cdot10^{-5}$~g and m$_2 = 1$~g, and mark the time at which the peak of the mass distribution reaches a mass corresponding to one of the check points.

\begin{figure}
   \centering
   \includegraphics[width=\hsize]{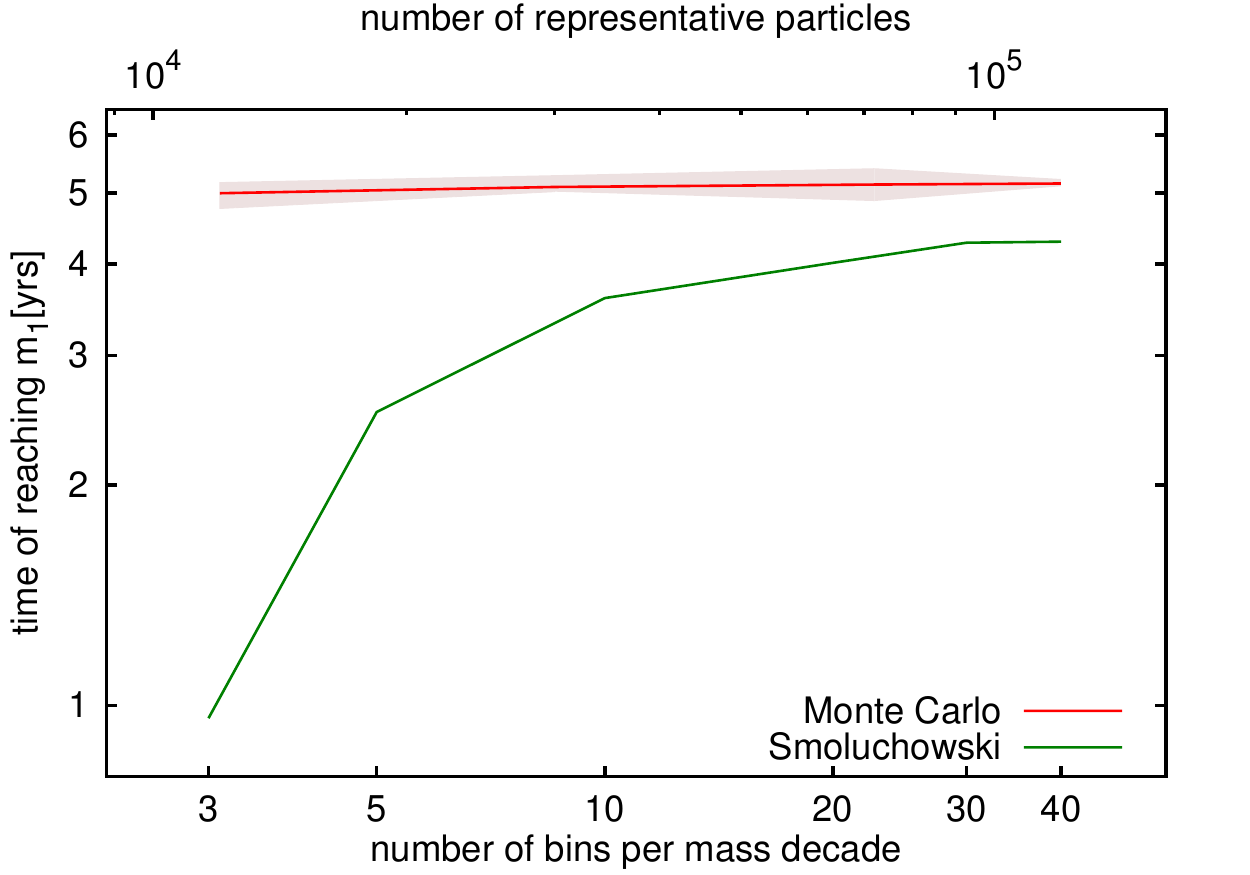}
      \caption{Time at which the peak of mass distribution reaches the mass m$_1 = 2\cdot10^{-5}$~g for both the Smoluchowski and Monte Carlo code. The result changes with mass resolution, given by the number of bins per mass decade in the case of Smoluchowski code and the number of particles used in the Monte Carlo code. The scatter of results obtained in different runs with the same number of particles in the Monte Carlo code is marked by the shaded region around an averaged dependence. The Monte Carlo approach does not exhibit a strong resolution dependence. In contrast, the Smoluchowski algorithm overestimates the growth rate by a factor of few when we do not use enough mass bins.}
         \label{fig:m1times}
\end{figure}

\begin{figure}
   \centering
   \includegraphics[width=\hsize]{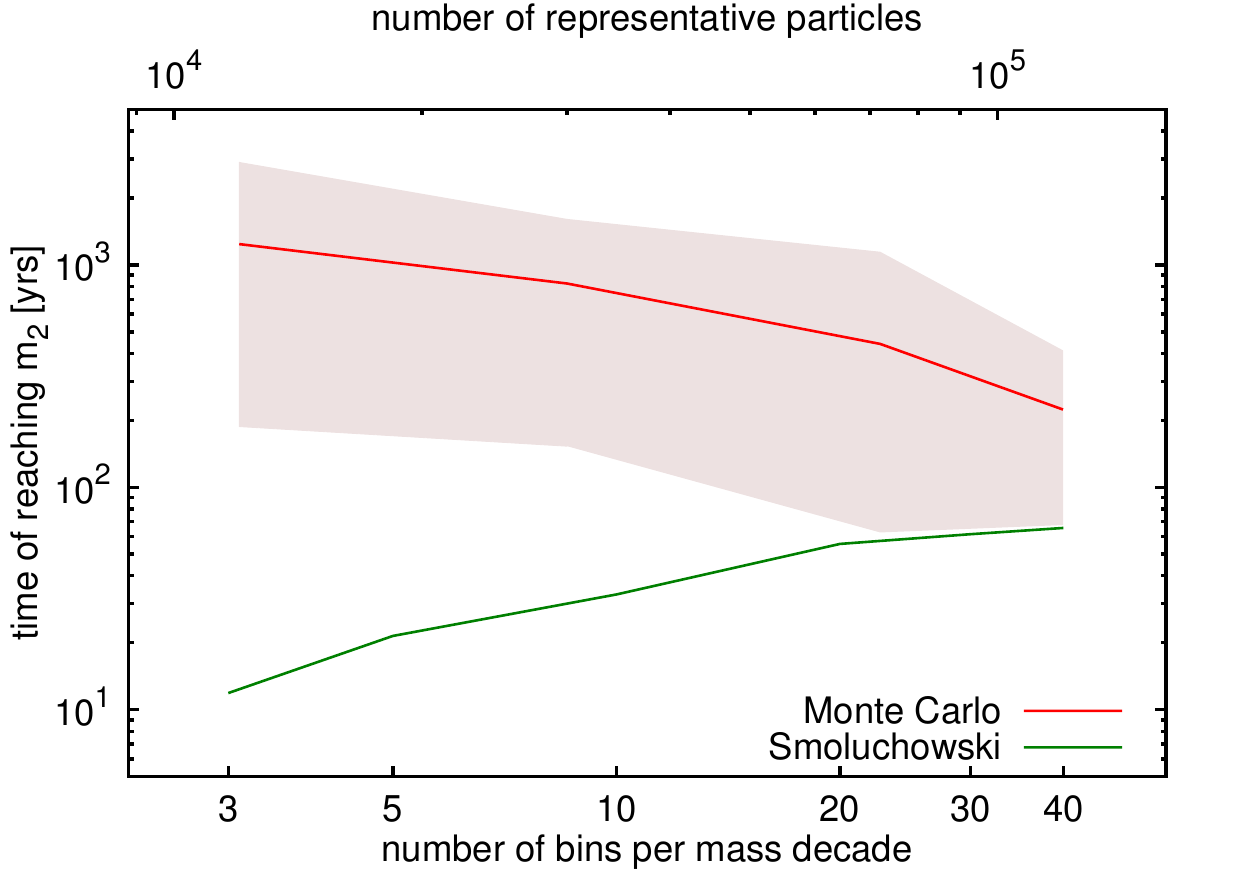}
      \caption{Figure analogical to Fig.\ \ref{fig:m1times} but for the “check point” m$_2=1$~g. The resolution dependence of the Smoluchowski method is the same as for m$_1$, but it changes significantly in the case of the Monte Carlo code. The Monte Carlo code tends to underestimate the breakthrough possibility when used with a low number of particles due to the limited dynamic mass range.}
         \label{fig:m2times}
\end{figure}

Figures \ref{fig:m1times} and \ref{fig:m2times} show the results obtained for m$_1$ and m$_2$, respectively. We show the times of crossing the “check points” as a function of the number of bins used in the Smoluchowski code and a number of representative particles used in the Monte Carlo method. We note that the scaling of x-axes of these figures is arbitrary, as there is no method to directly connect the number of bins in the Smoluchowski method to the number of representative particles in the Monte Carlo method.

Mass m$_1$ is located before the fragmentation barrier. This part of the evolution corresponds to a standard growth scenario that does not include the "lucky" breakthrough possibility and is well resolved by both of the codes. In the Monte Carlo code case, the time of crossing m$_1$ depends very weakly on the number of particles used. The difference between individual runs, as marked by the shaded region in Fig. \ref{fig:m1times}, is also very low. The Smoluchowski code exhibits much stronger resolution dependence. In the case of our lowest resolution of 3 bins per mass decade, the growth is more than four times faster than with 40 bins. The results obtained with resolution of 40 bins per mass decade converge to a value that is consistent with the one given by the Monte Carlo code. The critical resolution agrees with findings of \citet{2000Icar..143...74L} and \citet{2009ApJ...707.1247O}.


The reason for the high diffusion of the Smoluchowski method is the linearity of the algorithm described in Eq.\ \ref{epsilon}, which is a necessity for solvers that employ implicit integration schemes to overcome the numerical stiffness of the equations (see \citealt{2008A&A...480..859B} for further details). Explicit solvers, on the other hand, are capable of implementing higher order mass distribution schemes, which lowers the numerical diffusion. We find that steady states that arise when a bouncing or fragmentation barrier is met and that have no possibility of breaking through are significantly less dependent on mass resolution. Thus, the conclusions of most of the papers that did not include breakthrough are not affected, even if a lower resolution was used.

As can be seen in the Fig.\ \ref{fig:m2times}, the situation changes significantly in the case of the second check point m$_2$. This point is located beyond the fragmentation barrier and breakthrough point. Due to limited dynamic mass range, the resolution dependence of the Monte Carlo code is much stronger. The dispersion of results obtained in individual runs is much higher and can reach even an order of magnitude. Generally, the less particles we use, the longer we have to wait for the breakthrough. In the case of the Smoluchowski code, the resolution dependence is the same as for m$_1$, meaning that the lower resolution we use, the faster the breakthrough occurs. This is consistent with the findings of \citet{2013ApJ...764..146G}. The results obtained with both of the codes roughly converge for our highest accuracy. Both of the methods become computationally inefficient when used with even higher resolutions.

The different resolution dependence seen in the Fig.\ \ref{fig:m2times} is a result of a fundamental difference between the two approaches. In a real protoplanetary disk, the number of physical particles is so high that even when the breakthrough probability is low, some particles will be able to be "lucky" and overcome the fragmentation barrier quickly. However, the number of representative particles is restricted in the Monte Carlo code and the the breakthrough probability is additionally reduced. In contrast, the breakthrough is resolved much easier because the Smoluchowski code deals with number densities instead of discrete particles. However, this introduces another problem, which is discussed in the following section.

\subsection{Breakthrough probability}
\label{subsec:breakposs}

The "lucky growth" scenario has introduced a new issue in how the numerical diffusion affects the global dust evolution in the Smoluchowski method.
In this section, we discuss this issue and introduce a way to limit its effect by including a modulation function to the coagulation algorithm that suppresses the interactions with mass bins containing unrealistically low particle numbers.

As discussed above, when velocity distributions are introduced, the collision barriers are naturally smeared out. Because a particle that is more massive than the grains at the mass distribution peak must be "lucky" and has to grow by only interacting with other particles in low-velocity collisions, it becomes necessary to accurately resolve the high-mass tail of the distribution. Otherwise, if not all sticking events are resolved properly, the slope of the tail becomes incorrect, creating artificially large mass ratios between the luckiest grains and those in the peak. 

As an example of this, we can consider the extreme case where the entire high-mass tail is represented by a single mass bin $m_i$. If two grains in the peak undergo a single sticking event, forming a particle of mass $m \ll m_{i+1}$, some mass will still be put into the mass bin ${i+1}$, even though the particles would need to undergo several consecutive sticking events to reach a mass $m_{i+1}$ in reality. Such a badly resolved large-particle tail could cause an artificial breakthrough of growth barriers, as unrealistically large particles can form that continue to grow by the sweep-up process where no such particles would form in a better resolved case.

\begin{figure}
   \centering
   \includegraphics[width=\hsize]{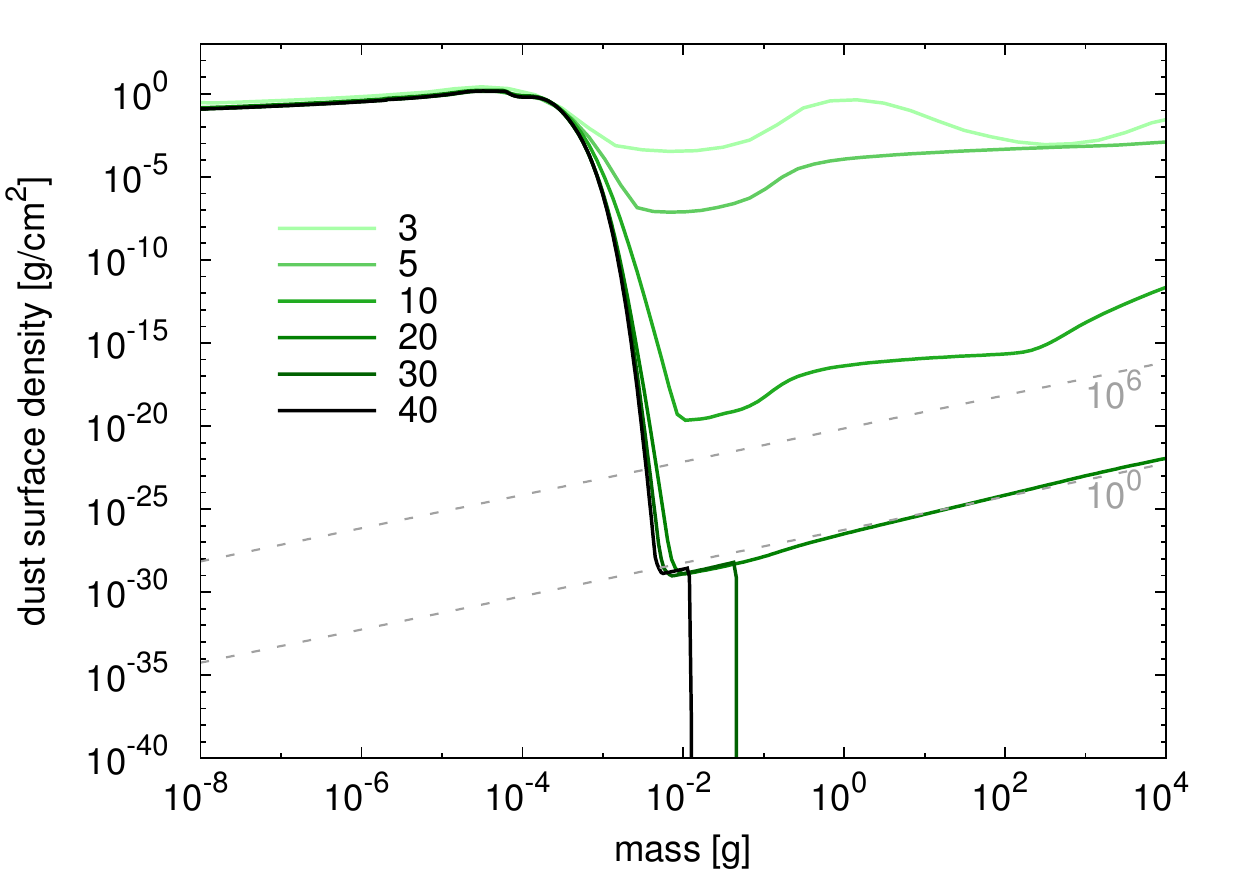}
      \caption{The effect of mass resolution in the case of SF$+$MT setup with a critical mass ratio of 500 for the Smoluchowski code. The resolutions included a range from 3 to 40, and the mass distribution is shown after 100 yrs of evolution. The point of breakthrough is very sensitive to the resolution used, and the breakthrough happens only in the low resolution cases, which is clearly nonphysical. The dashed lines mark the density corresponding to 1 and $10^6$ particles of a given mass in an annulus of 0.1 AU width.}
         \label{fig:Smoluchresol}
\end{figure}

To show this issue clearly, we perform additional simulations with the Smoluchowski code using a critical mass transfer ratio, that is the mass ratio above which a fragmenting collision leads to mass transfer (see Sect.\ \ref{sub:comp}), of 500. The point of breakthrough then occurs at a very low dust density, which is impossible to resolve with our implementation of the Monte Carlo method, as the particles that should break through would involve less mass than the mass of single swarm in our simulation (see also Sect.\ \ref{sub:model}). We have therefore performed a resolution study with the Smoluchowski code only.
Simulations were run with resolutions between 3 and 40 bins per decade of mass with the results shown in Fig.~\ref{fig:Smoluchresol}. As can be seen, the SF$+$MT case is extremely sensitive to the mass resolution. Between the highest and the lowest resolutions, the point of breakthrough differs by more than 25 orders of magnitude in surface density. For resolutions above 30 bins per decade, the point of breakthrough would have occurred at a density lower than the density corresponding to 1 particle in an annulus of 0.1 AU width.

To correctly simulate the lowest dust densities, it becomes necessary to include a modulation function, $f_{\rm mod}$, that limits the interactions between mass bins that have particle numbers that are too low. The reasoning behind this is the following. For breakthrough to occur, we need at least 1 real particle within the simulation domain that is large enough to trigger a sweep-up. Because the Smoluchowski code deals with number densities in a single point in space, the limiting particle density becomes somewhat arbitrary and relies on choosing a reasonable physical domain size.

The modulation function works in a similar way to the active bin method introduced by \citet{2000Icar..143...74L}. Instead of speeding up the code by deactivating low-populated mass bins; however, it is a continuous method that prevents growth of bins with unrealistically low particle numbers. The collision frequency between particle species $i$ and $j$ can then be written as
\begin{equation}\label{intrates}
	f_{ij} = n_i n_j K_{ij} \cdot f_{\rm mod},
\end{equation}
where $n_i$ and $n_j$ are the number densities, $K_{ij}$ the coagulation kernel. We choose $f_{\rm mod} = \exp{\left( -1/N_i -1/N_j \right)}$, where $N_i$ and $N_j$ are numbers of particles in a given surface of the disk, which we take to be an 0.1~AU wide annulus at 1~AU. The result of this is that mass is allowed to be put into a mass bin with a nonunity amount of particles by coagulation, for example, but the mass inside is unable to coagulate further until the bin contains sufficient mass.

For our setup, the modulation function $f_{\rm mod}$ suppresses the breakthrough in the case of resolution larger than 30 bins per mass decade. The breakthrough that occurs in the runs with lower resolution is a result of the numerical diffusion introduced by the mass distribution algorithm (Eq.\ \ref{epsilon}), which is discussed in the Sect.~\ref{sub:model}. This is seen by the sharp cut-off at masses between $10^{-2}-10^{-1}$ g for the highest resolutions.
We stress that this particular resolution dependency varies very strongly between setups, and it is necessary to confirm numerical convergence individually.

We want to note that even a high resolution run can lead to a nonphysical breakthrough if $f_{\rm mod}$ is not included. In such a case, breakthrough would initially occur at densities corresponding to less than one particle in an annulus of 0.1 AU width, but the density of the high-mass tail could increase over the threshold density during further evolution. Thus, the combination of sufficient numerical resolution and the modulation function is necessary to ultimately confirm the possibility of forming planetesimals in the breakthrough scenario under given conditions.

In this section, we clarified the issue of too crude mass resolutions found by \citet{2012A&A...548C...1W}. Although the lower resolutions commonly used in previous papers might work well for less extreme cases, which did not include the possibility of breakthrough, an artificial growth might occur for problems of the kind discussed here. Thus, we stress that careful convergence tests are necessary to confirm breakthrough possibility. We find that the convergence depends strongly on the strength of the growth barrier, and therefore, a separate convergence study is needed for each setup.

As was shown in the previous section, the Monte Carlo code underestimates the breakthrough chance with low particle numbers in contrast to the Smoluchowski code, and, thus, the time at which breakthrough happens increases. If the Monte Carlo code is used with not sufficient number of particles, the breakthrough might be completely suppressed. This happens when the mass that should be involved in the breakthrough is lower than $M_{\rm{swarm}}$, the mass represented by a single swarm, that limits the dynamic range of the Monte Carlo method (see Sect.~\ref{sub:model}). The minimum number of representative particles required is dependent on the breakthrough mechanism. Here, the breakthrough is driven by distribution in the impact velocities, which smear out the fragmentation barrier by changing the slope of the mass distribution high-mass end. The breakthrough is only possible if the largest particles are more massive than the particles in the mass distribution peak by a factor defined by the critical mass ratio. Thus, the number of breakthrough particles is defined by the mass distribution function slope and the value of the critical mass transfer ratio. The higher the slope and the critical mass ratio, the lower number of particles can break through. However, it was shown by \citet{2013A&A...556A..37D} that the breakthrough can also be driven by radial mixing of dust aggregates between regions of different grain sizes. In such case, the number of breakthrough particles follows different constraints than here.

\section{Discussion and conclusions}\label{sub:last}

The issue of numerical resolution of the Smoluchowski code has been discussed for a long time in the context of dust coagulation in protoplanetary disks \citep{1990Icar...83..205O, 1990Icar...88..336W,2000Icar..143...74L}. These codes used different algorithms that do not necessarily result in similar resolution dependencies as the implicit integration scheme with the linear mass distribution algorithm we use.

In this paper, we extended the prior studies by implementing a possibility of breaking through growth barriers with impact velocity distributions and including a direct comparison to the Monte Carlo algorithm with representative particle approach. The two methods have never before been explicitly compared. Our work showed that the two methods give consistent results when applied to usual coagulation problems. However, we find that modeling of the recently discovered planetesimal formation via "lucky growth" is much more challenging. Although the results obtained with the two methods converge for sufficient resolution, the approaches are fundamentally different and their limitations have to be realized when performing scientific models.

In agreement with previous studies, we find that simulations with the Smoluchowski code require a sufficiently high mass resolution to avoid an artificial speed-up of the growth rate. This problem arises from numerical diffusion and our implementation of the algorithm, which is required to describe how the resulting mass is distributed after a collision event. Additionally, we show that the introduction of the modulation function that prevents interactions between mass bins containing less than one physical particle is necessary to study the dust coagulation at low number densities. In the Sect.~\ref{subsec:breakposs}, we have shown that the numerical issues can change both the quantitative and the qualitative result in the case of the breakthrough scenario.

The Monte Carlo approach used to study the breakthrough scenario, in which only a few "lucky" particles break through the growth barriers, results in a high noise. In the presented tests, the individual runs show times of breakthrough that are orders of magnitude different and their averaged value depends on mass resolution. However, contrary to the Smoluchowski approach, the Monte Carlo approach does not suffer a strong resolution dependence when the dust aggregates grow with a relatively narrow size distribution, which usually is the case for small aggregate evolution when no breakthrough or runaway growth are possible. The Monte Carlo methods are generally computationally expensive, and they require numerous runs with different random seeds to reduce the noise.

The convergence of each of the methods can be very different for every setup. We cannot present a general recipe for the minimum resolution required to study dust growth, because this can vary enormously from case to case. Thus, it is important to run resolution tests for every new physical model until convergence of results is obtained.

Since we find that the numerical convergence is sensitive to the collision model parameters, it is important to use realistic values; however, these are poorly constrained, as the amount of data we have from laboratory experiments is restricted, and it is still not necessarily reproduced by direct numerical simulations.
Laboratory experiments show that bouncing collisions start to occur at velocities between 0.1 and 10 cm~s$^{-1}$ \citep{2013Icar..225...75K}, while numerical simulations claim that such collisions only rarely occur, if ever \citep{2011ApJ...737...36W, 2013A&A...551A..65S}. Fragmentation is also hotly discussed, as laboratory experiments find fragmentation to occur at velocities as low as a few cm~s$^{-1}$ between 5 cm grains \citep{2012ApJ...758...35S} and up to a few m~s$^{-1}$ for mm-sized grains \citep{Lammel:mLrgVib9}. Numerical simulations, on the other hand, predict significantly higher threshold velocities for fragmentation, ranging between 1 and 12 m~s$^{-1}$ for 6 to 10 cm-sized grains \citep{2011ApJ...737...36W, 2013MNRAS.435.2371M}. The value of the fragmentation threshold velocity determines the maximum size of particles that we are able to obtain before the breakthrough happens. Mass transfer experiments are even more uncertain with the mass transfer efficiency that ranges between 0 and 60\% \citep{2009MNRAS.393.1584T, 2010ApJ...725.1242K, 2011ApJ...736...34B,2013A&A...559A.123M}, and the critical mass ratio, which is in principle unexplored in the laboratory and only for small ratios numerically \citep{2013MNRAS.435.2371M,2013A&A...559A..62W}. In all of these cases, additional material and collisional properties, such as porosity, composition, structure and impact angle also greatly influence the outcome, which adds to the uncertainty. As discussed in this work, the critical mass transfer ratio might need to be significantly lower than estimated in prior studies, due to the need for a relatively high numerical resolution of Smoluchowski solvers to accurately represent the high-mass tail.

The collision models used in sweep-up modeling attempt to simplify the very complex physics of collisions between dust agglomerates. There have been a few attempts at more rigorous models \citep{2010A&A...513A..57Z, 2012A&A...540A..73W} to consolidate the recent progress in laboratory and numerical collision experiments \citep{2007ApJ...661..320W,2008ARA&A..46...21B,2010A&A...513A..56G}, but these are still not necessarily more correct than the simple models that narrow the modeling down to only a few key parameters \citep{2012A&A...544L..16W, 2013ApJ...764..146G, 2013ApJ...774L...4M, 2013A&A...556A..37D}. As the critical mass ratio plays the crucial role in determining if the breakthrough is possible, restricting its realistic values should be a priority for the future laboratory studies.

We find the breakthrough scenario to be more sensitive to resolution issues than other problems we have tested, when the dust growth is utterly stopped by bouncing or fragmentation. At the same time, this  scenario is of particular importance for the current planetesimal formation theory.
Regardless of the method used, modeling of the "lucky growth" requires extreme computational force, and restricting the resolution to make the models more efficient can lead to serious numerical artifacts: the nonphysical breakthrough in the Smoluchowski approach case and lack of breakthrough in the case of the Monte Carlo approach.
 
To conclude, we list features that characterize the two main coagulation methods. One should keep these in mind when deciding which method to use for a particular scientific application.

The Smoluchowski equation solver with implicit integration scheme
\begin{itemize}
\item is capable of simulating dust evolution over long timescales (even at high resolutions, 0D simulations are finished within minutes),
\item resolves equilibrium states well, as the implicit integration scheme allows for very large time-steps once the solution approaches steady-state \citep{2010A&A...513A..79B},
\item has a very high dynamical range that allows phenomena involving a single physical particle of tiny mass (compared to the total dust mass) to be resolved, which makes it ideal for breakthrough studies \citep{2012A&A...540A..73W}, and to produce synthetic observations, as the opacity is dominated by small particles, which may contain only a low fraction of the dust mass,
\item is slowed by a factor of $\mathcal{O}(n^3)$, where $n$ is the number of mass bins, for each additional dust property beyond mass \citep{2012A&A...540A..73W}, although numerical tricks exist to circumvent this (see, e.g., \citealt{2009ApJ...707.1247O}),
\item suffers from high numerical diffusion that affects both growth timescale and breakthrough likelihood. The growth timescale can be benchmarked against the analytical kernels (e.g.\ \citealt{1990Icar...83..205O}), but it depends strongly on the strength of the barrier in the breakthrough case.
\end{itemize}
The Monte Carlo coagulation algorithm with equal-mass representative particles
\begin{itemize}
\item makes it easy to implement additional particle properties, such as porosity \citep{2008A&A...489..931Z, 2010A&A...513A..57Z}, because the computation time does not depend significantly on the number of properties that are evolved, but only on the number of collisions performed,
\item it is straightforward to develop it to further spatial dimensions \citep{2011A&A...534A..73Z, 2013A&A...556A..37D}, as the representative particles can be treated as Lagrangian tracer bodies,
\item can be used along with hydrodynamic grid codes \citep{2012A&A...537A.125J},
\item experiences no numerical diffusion of the mass function in general, so there is no danger of encountering an artificial speed-up of the growth,
\item has difficulty resolving features that include low fraction of total mass, which makes it less useful in the case of breaking through the growth barriers or runaway growth modeling, although the algorithm can be developed to overcome this issue \citep{2008ApJ...684.1291O},
\item makes it hard to model evolution over long timescales in general, because it is impossible to use extremely long time-steps, as every collision needs to be resolved.
\end{itemize}

\begin{acknowledgements}
We thank the anonymous referee as well as Chris Ormel, Andras Zsom, Satoshi Okuzumi, Til Birnstiel, Sebastian Stammler and Sebastian Lorek for useful comments that helped to improve the manuscript.
J.D. was partially supported by the Innovation Fund FRONTIER of the Heidelberg University. F.W. was funded by the Deutsche Forschungsgemeinschaft within the Forschergruppe 759 “The Formation of Planets: The Critical First Growth Phase”.
J.D. would also like to acknowledge the use of the computing resources provided by bwGRiD (http:$\backslash\backslash$www.bw-grid.de), member of the German D-Grid initiative, funded by the Ministry for Education and Research (Bundesministerium f\"{u}r Bildung und Forschung) and the Ministry for Science, Research and Arts Baden-Wuerttemberg (Ministerium f\"{u}r Wissenschaft, Forschung und Kunst Baden-W\"{u}rttemberg)
\end{acknowledgements}
\bibliographystyle{aa}
\bibliography{breakthrough_comparison.bib}
\end{document}